\documentclass[a4paper]{spie}  
\usepackage[]{graphicx}
\usepackage{txfonts}
\title{Cold performance tests of blocked-impurity-band Si:As detectors developed for {\textsc{Darwin}}} 

\author{Stephan M. Birkmann\supit{a}, Jutta Stegmaier\supit{a}, Ulrich Gr\"ozinger\supit{a}, Oliver Krause\supit{a}, Tim Souverijns\supit{b}, Jan Putzeys\supit{b}, Deniz Sabuncuoglu Tezcan\supit{b}, Koen De Munck\supit{b}, Paolo Fiorini\supit{b}, Kiki Minoglou\supit{b}, Patrick Merken\supit{b}, Chris Van Hoof\supit{b}, Piet De Moor\supit{b}
\skiplinehalf
\supit{a}Max-Planck-Institut f\"ur Astronomie, K\"onigstuhl 17, D-69117 Heidelberg, Germany; \\
\supit{b}IMEC, Kapeldreef 75, B-3001 Leuven, Belgium
}

\authorinfo{Send correspondence to S.M.B., E-mail: birkmann@mpia.de}

\begin{document} 
  \maketitle 

\begin{abstract}
We report first results of laboratory tests of Si:As blocked-impurity-band (BIB) mid-infrared (4 to 28$\,\mu$m) detectors developed by IMEC. These prototypes feature 88 pixels hybridized on an integrated cryogenic readout electronics (CRE). They were developed as part of a technology demonstration program for the future \textsc{Darwin} mission. In order to be able to separate detector and readout effects, a custom build TIA circuitry was used to characterize additional single pixel detectors. We used a newly designed test setup at the MPIA to determine the relative spectral response, the quantum efficiency, and the dark current. All these properties were measured as a function of operating temperature and detector bias. In addition the effects of ionizing radiation on the detector were studied. For determining the relative spectral response we used a dual-grating monochromator and a bolometer with known response that was operated in parallel to the Si:As detectors. The quantum efficiency was measured by using a custom-build high-precision vacuum black body together with cold ($T \sim 4\,$K) filters of known (measured) transmission.
\end{abstract}



\section{INTRODUCTION}
\label{sec:intro}  

In this paper we present first results of Si:As block-impurity-band detector that were developed by the Belgium research institute IMEC as part of a technology study for the European Space Agency (ESA) in the framework of the future \textsc{Darwin} mission\cite{fridlund:62680R}. This mission targets the discovery and classification of Earth like planets around nearby stars. \textsc{Darwin} will be a flotilla of 4 to 5 satellites\cite{wallner:626827} and will utilize nulling interferometry for canceling out the light of the host star in order to be able to analyze the chemical signatures of the much fainter planets.

Such a mission requires high performance detectors in the wavelength range of 6 to 18\,$\mu$m and preferably beyond, providing low dark currents and high quantum efficiency. Blocked-impurity-band detectors like those based on Si:As can deliver the needed performance and have additional advantages when compared to bulk detectors, like better signal stability and radiation hardness\cite{2002dlus.book.....R}.

In the next section we will describe the tested detectors, followed by a description of the test setup in section~\ref{sec:testsetup}, and reporting the results in section~\ref{sec:results}.

\section{TEST SPECIES}
\label{sec:species}

The details of detector fabrication and the properties of their readout electronics were presented by Tezcan et al.~(2007)\cite{tezcan:66600R}. The detector is fabricated on a high resistive silicon substrate for backside illumination. The buried contact and both the active and blocking layers are deposited by epitaxy. Access to the buried contact is provided by anisotropic silicon etch of V-grooves.

\begin{figure}[tb]
   \begin{center}
   \includegraphics[width=0.49\textwidth]{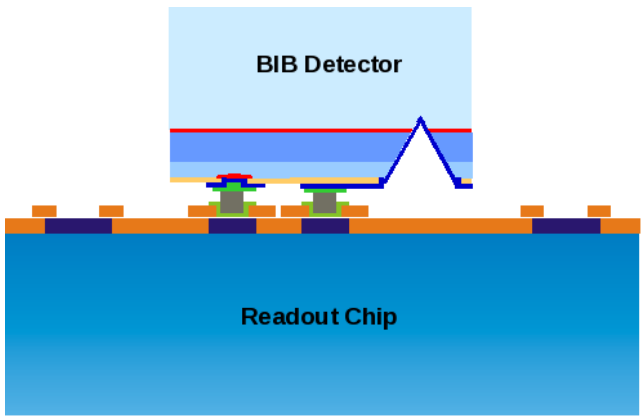} \hspace{\stretch{1}}
   \includegraphics[width=0.49\textwidth]{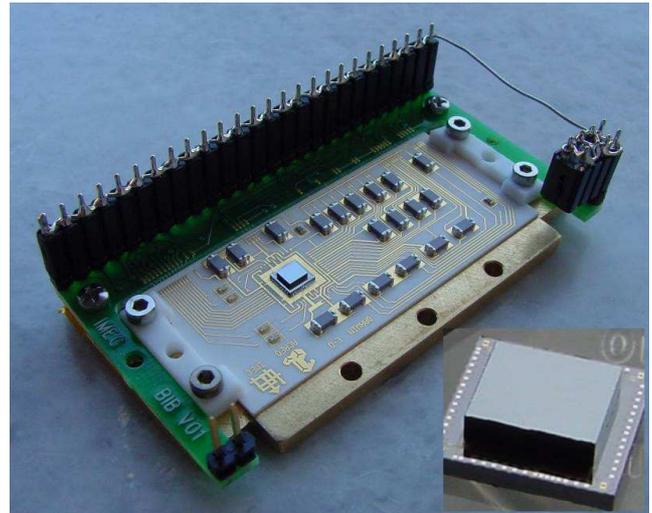}
   \caption{ \label{fig:cre} Left: Schematic view of the detector/readout interface of the BIB prototypes. The linear detector array is flip-chip mounted directly on the cryogenic readout electronics using indium bumps.\newline Right: The front end electronic with the CRE and detector. In the lower right a magnified view of the CRE plus detector is displayed.}
   \end{center}
\end{figure} 

The cold readout electronics (CRE) is composed of an AC coupled integrating amplifier with capacitive feedback (CTIA). It also features a sample and hold stage and an output buffer with a multiplexing unit. This readout is a direct heritage of the \textsc{Herschel/Pacs}\cite{poglitsch:62650B} Ge:Ga readout electronics developed by IMEC\cite{merken:627516}. It was adapted to use different feedback capacities and to operate a linear array of up to 88 pixels ($30\,\mathrm{\mu m}$ pitch). Detector arrays are coupled to the CRE by indium bumps using flip-chip technology (see Fig.~\ref{fig:cre}, left). The CRE and the detector are mounted on a alumina substrate providing the mechanical and electrical interface, the so called front end electronic (FEE) (see Fig.~\ref{fig:cre}, right).

\begin{figure}[tb]
   \begin{center}
   \includegraphics[width=0.4\textwidth]{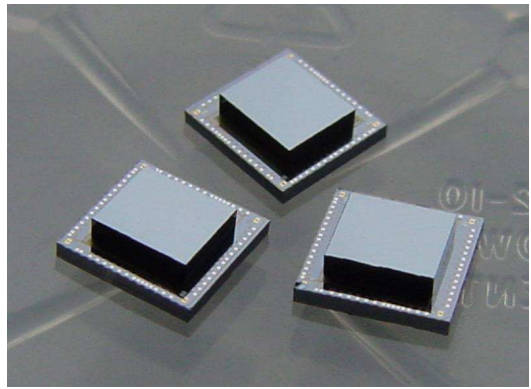}
   \caption{ \label{fig:testdiode} Microscopic view of the top surface of test detector \#1, with the metalization completely covering the implant. In the middle the groove for accessing the bottom contact can be seen. The larger metalized pads on the right are for bonding purposes.}
   \end{center}
\end{figure}

While well suited for the relatively high background in \textsc{Pacs}, these CREs have a leakage current too high to be used for dark current measurements of the small area CRE mounted BIB detectors. In order to study the detector properties more independent from the readout, IMEC fabricated different large area single pixel test detectors mounted in a ceramic DIL package. These devices can be contacted and readout using a custom electronics. The differences between the detectors are the layout of the top contact - including the metalization - and the effective active area (see Fig.~\ref{fig:testdiode}). A summary of the characteristics of the three tested devices is given in table~\ref{table:devices}.

\begin{table}[htb]
 \begin{center}
\caption{ \label{table:devices} The three test detectors and their parameters.}
  \begin{tabular}{c c c}\hline\hline\\[-6pt]
   \textbf{Device number}&\textbf{Description}&\textbf{Active area [cm$^2$]}\\[4pt]\hline
   & & \\[-6pt]
   \#1 & hidden top contact, metalization covering the implant & $5.21\times 10^{-3}$\\
   \#2 & shortened array of 88 linear pixels & $6.83\times 10^{-4}$\\
   \#3 & uncovered full top contact, metalization only at edge & $2.50\times 10^{-2}$\\[4pt]\hline\hline
  \end{tabular}
 \end{center}
\end{table}

\section{TEST SETUP}
\label{sec:testsetup}

Here we describe the test setups for (i) the absolute quantum efficiency / responsivity and dark current measurements and (ii) the relative spectral response measurements. Both setups share the same cryostat, an IRLabs two vessel bath cryostat with a 5 inch coldplate. The detector holder is made of oxygen free copper and located in the middle of the detector housing made of aluminum. The holder is separated from the base by means of four graphite rods glued with EcoBond 206. The graphite is used to achieve both a good thermal conduction between the cold plate and the detector during cool down and to limit the thermal conduction at low temperatures. This allows the detector temperature to be adjusted between $T_\mathrm{Det} = 4.5$\,K and $T_\mathrm{Det} = 15$\,K by using a Lakeshore 340 Temperature Controller with little power dissipation ($P_\mathrm{Heater} \leq 40\,\mathrm{\mu W}$ at $T_\mathrm{Det} = 6$\,K).

\subsection{Responsivity and Dark Current}

\begin{figure}[tb]
 \begin{center}
  \includegraphics[height=0.42\textwidth]{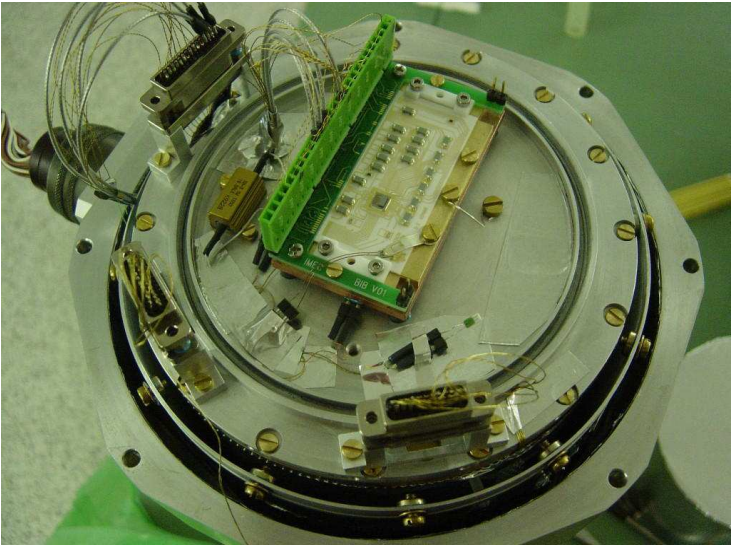} \hspace{\stretch{1}}
  \includegraphics[height=0.42\textwidth]{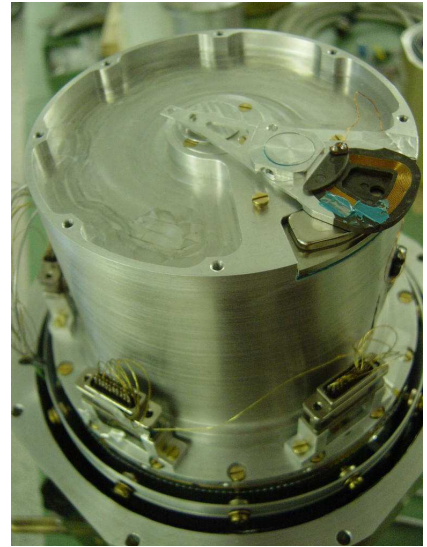}
   \caption{ \label{fig:setup} Left: The coldplate with a mounted FEE and detector.\newline Right: The detector housing with the pinhole and cold shutter. The cold bandpass filter is located beneath the pinhole (not visible).}
 \end{center}
\end{figure}

For the responsivity and dark current measurements the detector under test is placed into the copper holder (see Fig.~\ref{fig:setup}, left) and a light tight aluminum housing is put onto the base plate. The detector is illuminated through a pinhole with a diameter of $0.3\,$mm located at the top of the housing. The light also has to pass a narrow band filter of measured transmission with a bandpass centered at $\lambda_c = 11.3\,\mathrm{\mu m}$. Because the bandpass filter is thermally anchored to the LHe level of the test cryostat, the thermal emission of the filter is negligible.

As illuminating source We use a custom built black body simulator that is located inside the vacuum chamber of the cryostat. It is made of aluminum and has a double cone cavity with the internal surfaces being roughed and coated with IR-black paint. Together with the coating the double cone design yields a high emissivity of $E \geq 0.99$\cite{Bedford:76}. The temperature is measured and controlled using a calibrated Pt-100 sensor ($\Delta T_\mathrm{BB} = 0.25$\,K) and by heater wire wound around the black body, using a Lakeshore 340 Temperature Controller. The standard operating temperature of the black body is $T_\mathrm{BB} = 295$\,K.

The flux onto the detector is determined by the black body temperature and emissivity, the pinhole geometry, and the transmission of the bandpass filter. Because of the relatively warm black body the uncertainty of the flux $F$ due to temperature is small ($\Delta F/F (\Delta T) \leq 0.5$\%). The transmission of the cold bandpass filter was measured with an accuracy of $\pm 2$\%. Therefore the total flux uncertainty is governed by the diameter of the pinhole and its distance to the detector and is estimated to $\Delta F/F = 5$\%.

In order to perform differential measurements and estimate the detector dark current, the housing is also equipped with a cold shutter ($T \sim 10$\,K) that covers the pinhole (see Fig.~\ref{fig:setup}, right). The shutter has been checked for light leaks by performing designated measurements under absolute dark conditions, i.e.\ with the detector completely covered with aluminum tape.

\subsection{Relative Spectral Response}

In order to measure the relative spectral response in the range of $5\,\mathrm{\mu m} \leq \lambda \leq 18\,\mathrm{\mu m}$ we operate an IRLabs Si-bolometer with a flat response in parallel to the detector under test in the same cryostat. The IR radiation is provided by a laboratory black body and a nitrogen purged dual grating monochromator. The latter is also used to measure the transmission of infrared filters\cite{birkmann:62750S}.

\begin{figure}[tb]
 \begin{center}
  \includegraphics[width=0.7\textwidth]{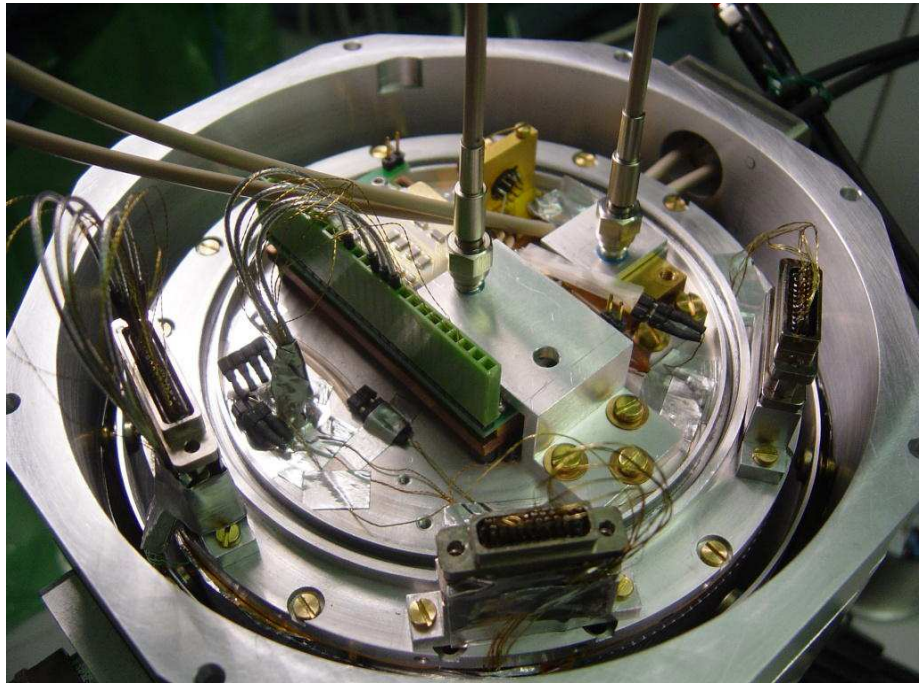}
   \caption{ \label{fig:fibers} The test setup for the relative spectral response measurements. Two fibers guide the light from the monochromator to the detector under test and the bolometer (used as reference).}
 \end{center}
\end{figure}

The simultaneous illumination of the two detectors in the Dewar with identical spectral content is accomplished by using two mid-infrared single mode fibers developed by ART Photonics that provide good transmission in the wavelength range from 4 to 18\,$\mu$m\cite{artiouchenko:295}. One end of the fibers is located at the exit slit of the monochromator where the light is coupled in. The fibers enter the cryostat through a vacuum tight interface and their other ends are directly mounted on top of the detector under test and the reference bolometer respectively (see Fig.~\ref{fig:fibers}).

\subsection{Readout Electronics}
\label{sec:readout}

For first tests of the linear arrays with the CRE flip-chip mounted we used the same warm readout electronics as for the PACS Ge:Ga detectors\cite{birkmann:437,stegmaier:62652H}. For the three single pixel test detectors the cold readout electronics is composed of an IRLabs transimpedance amplifier (TIA) with paired JFETS in conjunction with a cold feedback resistor ($8\times 10^7\,\Omega \leq R_\mathrm{F} \leq 1.3\times 10^9\,\Omega$) from ELTEC. The resistance of these resistors is voltage and temperature dependent and was measured beforehand. The warm electronics is build up using commercial integrated circuits. Two dual channel zero-drift operational amplifiers (LTC2051HV) are used for detector signal amplification and feedback and also for buffering the bias supply. The bias voltage is set via a computer using a low-noise, monotonic 16-bit DAC with I2C interface and a low-noise, low-drift voltage reference (ADR430).

The amplified output voltage $U_\mathrm{Out}$ is measured using a nano volt meter (Agilent 34420A) and a spectrum analyzer. Additionaly, two digital lock-in amplifiers (Signal Recovery Model 7225) are used for the relative spectral response measurements. The data is recorded and analyzed using a PC with Agilent Vee and IDL routines.

\section{RESULTS}
\label{sec:results}

In this chapter we present the results obtained with three large area single pixel test detectors using the TIA readout electronics described above.

\subsection{Dark Current}

The dark current of the three devices is measured in dependence of bias voltage $U_\mathrm{Bias}$ and detector temperature $T_\mathrm{Det}$, ranging from $0.1 \leq U_\mathrm{Bias} \leq 1.5$\,V and $4.5 \leq T_\mathrm{Det} \leq 15$\,K. In order to ensure absolute dark conditions the detectors are covered by aluminum tape, in addition to the closed cold shutter. The dark current is measured as
\begin{equation}
 I_\mathrm{Dark} = \frac{U_\mathrm{Out}\left(U_\mathrm{Bias}\right) - U_\mathrm{Out}\left(U_\mathrm{Bias} = 0\,\mathrm{V}\right)}{R_\mathrm{F}\left(U_\mathrm{Out}\right)},
\end{equation}
where $U_\mathrm{Out}$ is the output voltage of the readout electronics and $R_\mathrm{F}\left(U_\mathrm{Out}\right)$ the output voltage dependent resistance of the feedback resistor.

\begin{figure}[tb]
\begin{center}
\includegraphics[width=0.49\textwidth]{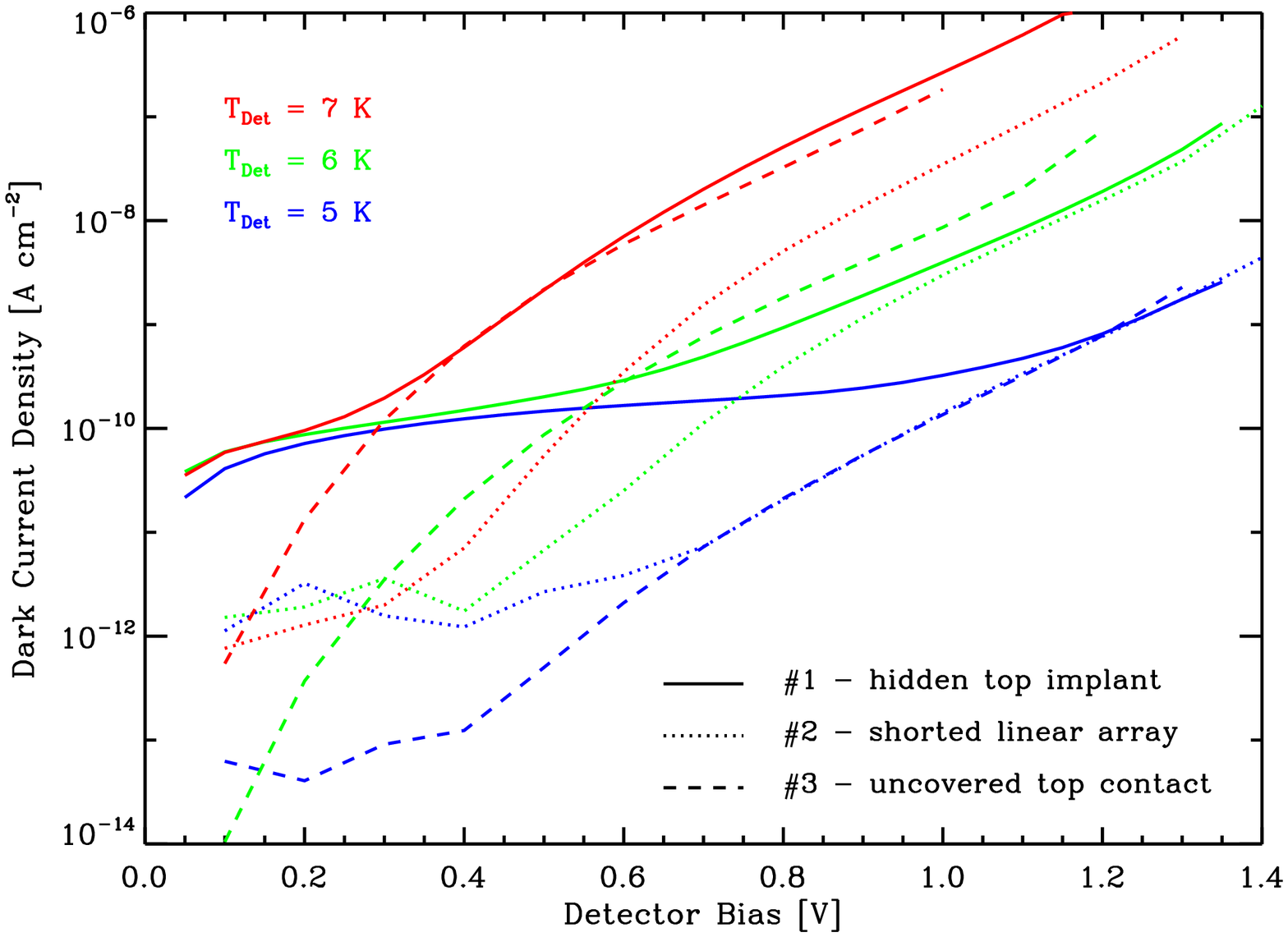} \hspace{\stretch{1}}
\includegraphics[width=0.49\textwidth]{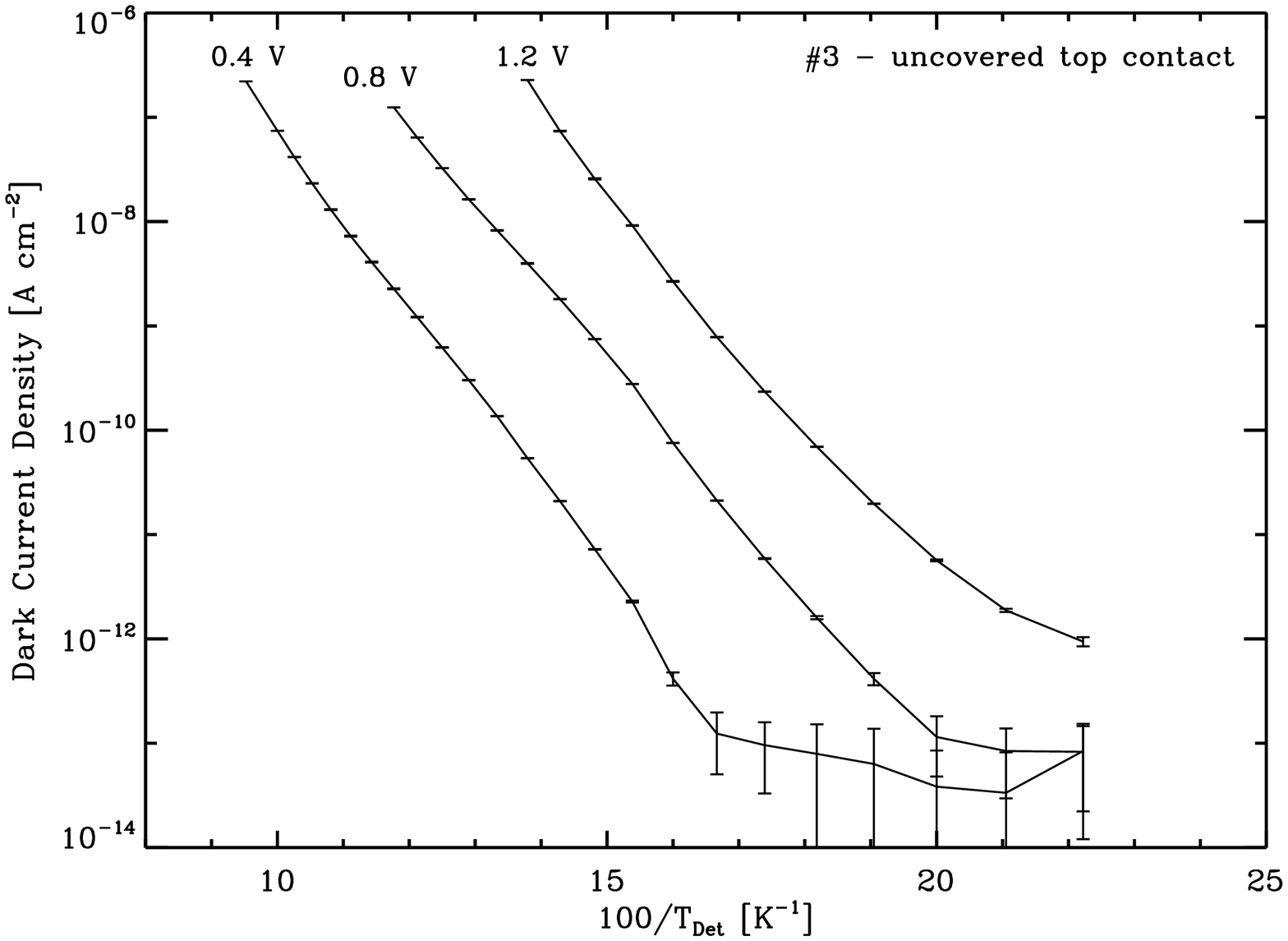}
\caption{\label{fig:dc}Left: The dark current density as measured for the three test devices. As expected, the dark current increases with detector temperature and bias voltage. Please note that detector \#1 was measured with a feedback resistor of smaller value, resulting in a higher measurement error for low currents.\newline Right: The dark current for test device \#3 in dependence of temperature ($4.5\,\mathrm{K} \leq T_\mathrm{Det} \leq 10.5\,\mathrm{K}$) for three different bias voltages.}
\end{center}
\end{figure}

The results for the three detectors are shown in Fig.~\ref{fig:dc}. In the left panel the dark current density (dark current normalized to the detector area) is displayed for three different detector temperatures and a range of bias voltages. Please note that for detector \#1 we used a different feedback resistor with smaller resistance ($R_\mathrm{F}\left(0\,\mathrm{V}\right) = 0.5 \times 10^9\,\Omega$ compared to $R_\mathrm{F}\left(0\,\mathrm{V}\right) = 1.3 \times 10^9\,\Omega$). Together with the smaller detector area this leads to a larger measurement uncertainty, especially affecting the low dark currents at low bias voltages. For $U_\mathrm{Bias} \geq 0.8$ the dark current density agrees quite well for the different detector types.

In the right panel of Fig.~\ref{fig:dc} the dark current density for detector \#3 is displayed for three different values of 
$U_\mathrm{Bias}$ in dependence of detector temperature. At low temperatures ($T_\mathrm{Det} \leq 5\,\mathrm{K}$) the dark current density is below $J_\mathrm{Dark} \leq 10^{-11}\,\mathrm{A\, cm^{-2}}$. For a pixel size of $30\,\mathrm{\mu m} \times 30\,\mathrm{\mu m}$ this would translate to a dark current of $I_\mathrm{Dark} \approx 500\,\mathrm{e\, s^{-1}}$. For moderate bias voltages ($U_\mathrm{Bias} = 0.8$\,V) this is further reduced to $I_\mathrm{Dark} \approx 5\,\mathrm{e\, s^{-1}}$.

\subsection{Activation Energy and Cut-off Wavelength}

From the Arrhenius-style plot in the right panel of Fig.~\ref{fig:dc} we estimate the activation energy via the formula
\begin{equation} \label{eq:dc}
 I_\mathrm{Dark} \propto \exp{\left(-\frac{E_\mathrm{c}-E_\mathrm{d}}{k T_\mathrm{Det}} \right)},
\end{equation}
where $k = 8.617\times 10^{-5}\,\mathrm{eV\, K^{-1}}$ is Boltzmann's constant, $E_\mathrm{c}$ the energy level of the conductor band, and $E_\mathrm{d}$ the energy level of the donators. The results for all three detectors agree, with some scatter when comparing different measurements for different bias voltages. As a mean value we find $E_\mathrm{c}-E_\mathrm{d} = 27 \pm 6\,\mathrm{meV}$.

In principle it is possible to estimate the cut-off wavelength of the detectors from the derived activation energy, however this requires the knowledge of the Fermi level $E_\mathrm{f}$. Because the latter is a function of detector temperature as well as donator and acceptor concentrations, this is not straightforward. At very low temperature the Fermi level is close to the donator level, therefore yielding an estimate of the cut-off wavelength $\lambda_\mathrm{co}$ via the formula
\begin{equation}
 \lambda_\mathrm{co} = \frac{1.24\,\mathrm{eV}}{E_\mathrm{g}}\,\mathrm{\mu m}.
\end{equation}
This yields a projected cutoff wavelength of $\lambda_\mathrm{co} = 46^{+13}_{-9}\,\mathrm{\mu m}$ which is significantly higher than the expected $\approx 28\,\mathrm{\mu m}$.

\begin{figure}[tb]
\begin{center}
\includegraphics[width=0.49\textwidth]{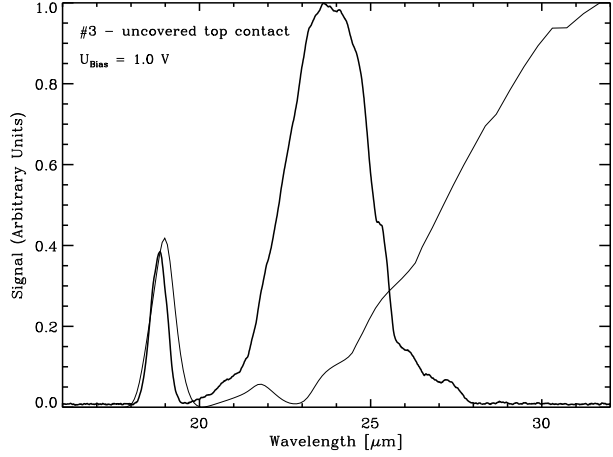} 
\caption{\label{fig:bandgap}The normalized detector signal (thick line) for the wavelength range $16\,\mathrm{\mu m} \leq \lambda \leq 32\,\mathrm{\mu m}$ when illuminated through a cut-on filter. The transmission of the filter is denoted by the thin line. Because there was no reference detector (bolometer) for this measurement, the BIB detector signal was not corrected neither for the grating efficiency, atmospheric transmission, nor the black body spectrum.}
\end{center}
\end{figure}

We therefore measure the cut-off wavelength directly using the monochromator and a cold cut-on filter of known transmission. The result is shown in Fig.~\ref{fig:bandgap}, indicating that the detectors become unresponsive for wavelengths $\lambda > 28\,\mathrm{\mu m}$. This suggests that the derivation of the cut-off wavelength for Si:As detectors by dark current measurements is rather inaccurate and biased towards longer wavelengths, as has been reported by others\cite{hogue:667809}.

\subsection{Responsivity and Quantum Efficiency}
\label{sec:resp}

We measure the responsivity of the test detectors by illuminating them with a known flux and performing differential measurements using the shutter. Therefore offset effects in the readout and the dark current are cancelled out. When the shutter is open the flux density on the detectors is $F = 1.89 \times 10^{-9}\,\mathrm{W\, cm^{-2}}$ at a wavelength of $\lambda = 11.3\,\mathrm{\mu m}$ (due to the used bandpass filter), corresponding to a photon flux density $Q = 1.08 \times 10^{11}\,\mathrm{photons\, s^{-1}\, cm^{-2}}$. The responsivity is calculated to
\begin{equation} \label{eq:resp}
 \Re = \frac{I_\mathrm{Det,open} - I_\mathrm{Det,closed}}{F \times A_\mathrm{Det}},
\end{equation}
where $I_\mathrm{Det,open}$ and $I_\mathrm{Det,closed}$ are the measured detector currents with open and closed shutter, respectively, and $A_\mathrm{Det}$ is the active detector area.

\begin{figure}[p]
 \begin{center}
  \includegraphics[height=0.3\textheight]{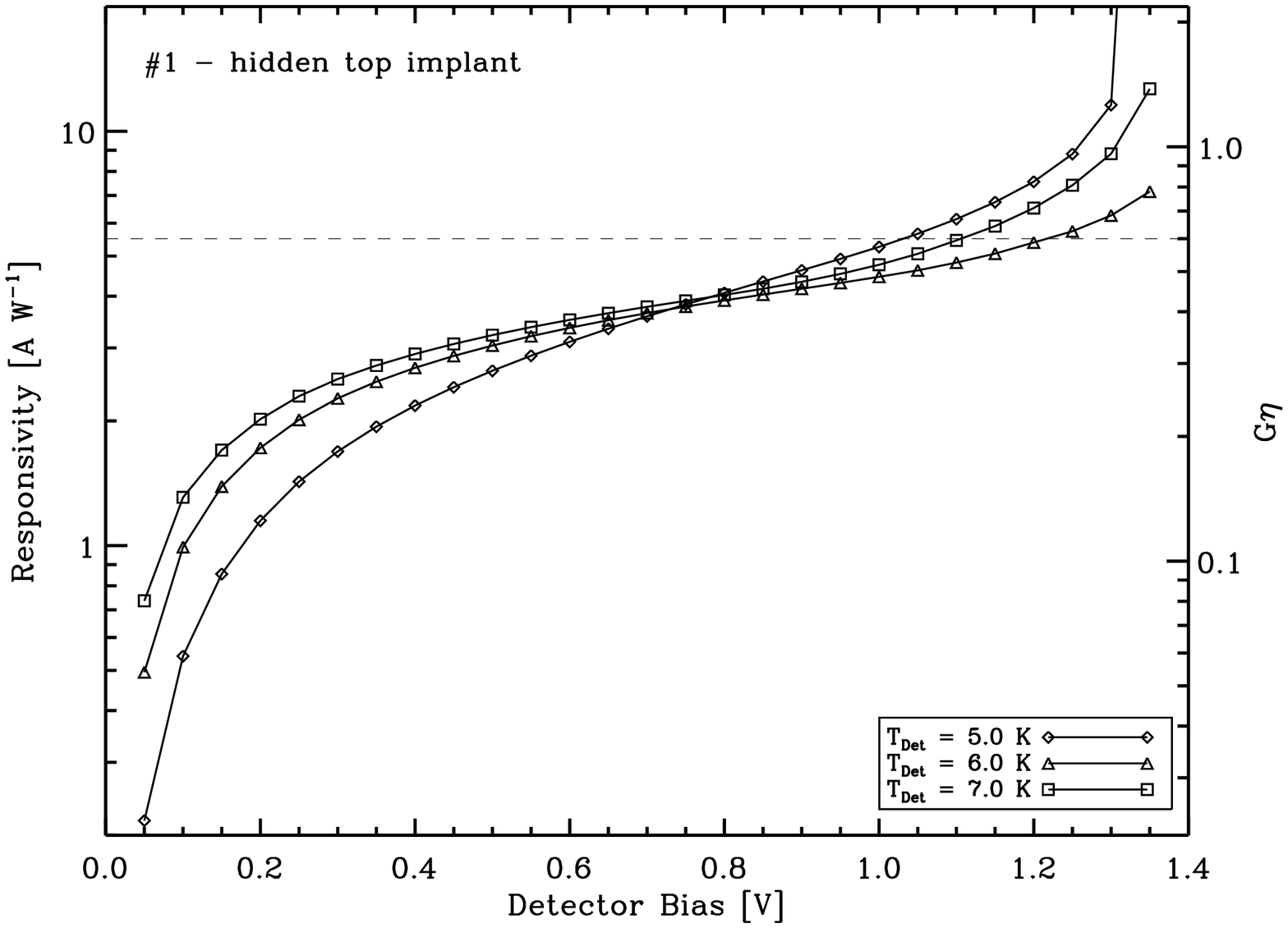}
  \includegraphics[height=0.3\textheight]{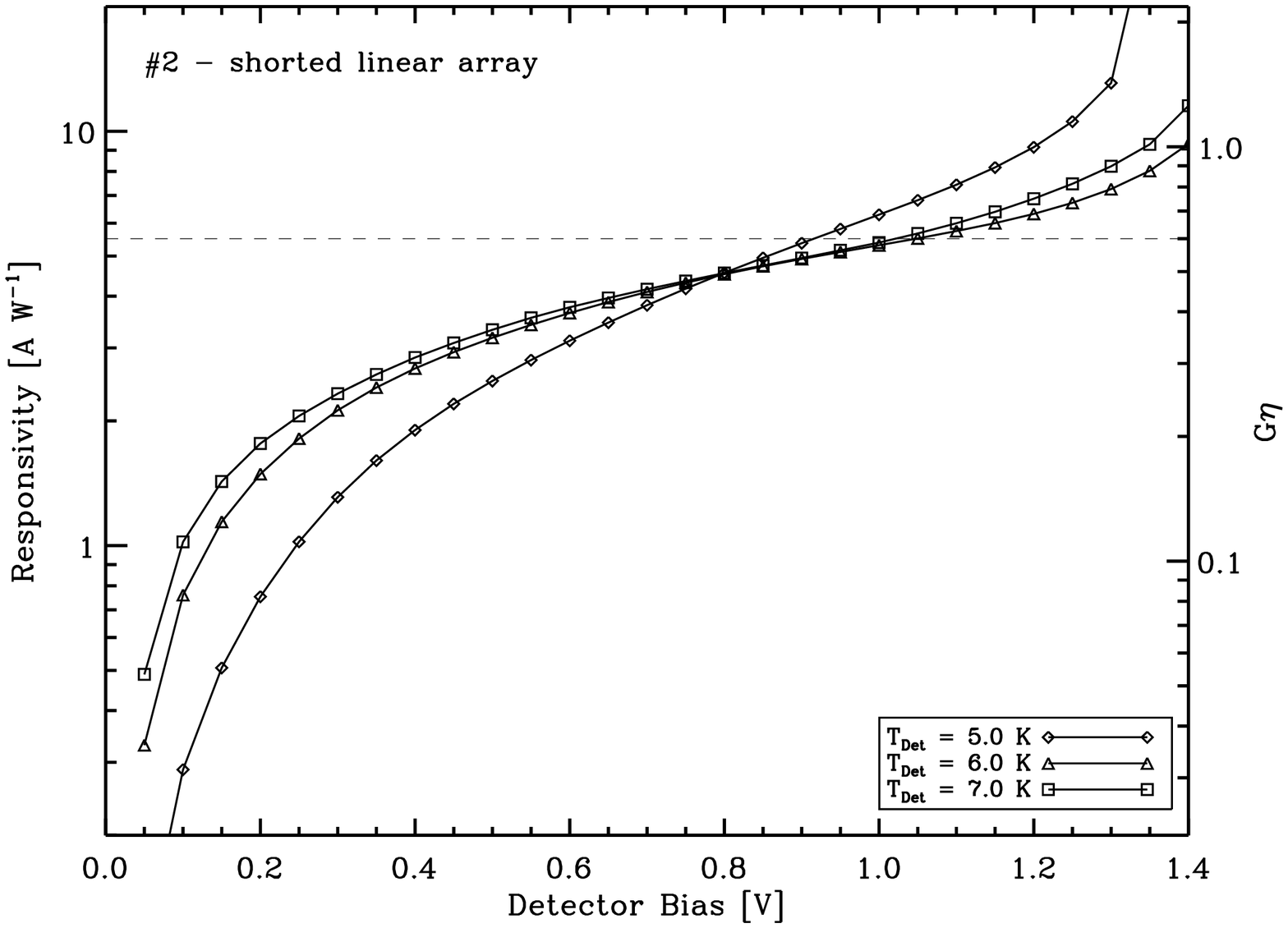}
  \includegraphics[height=0.3\textheight]{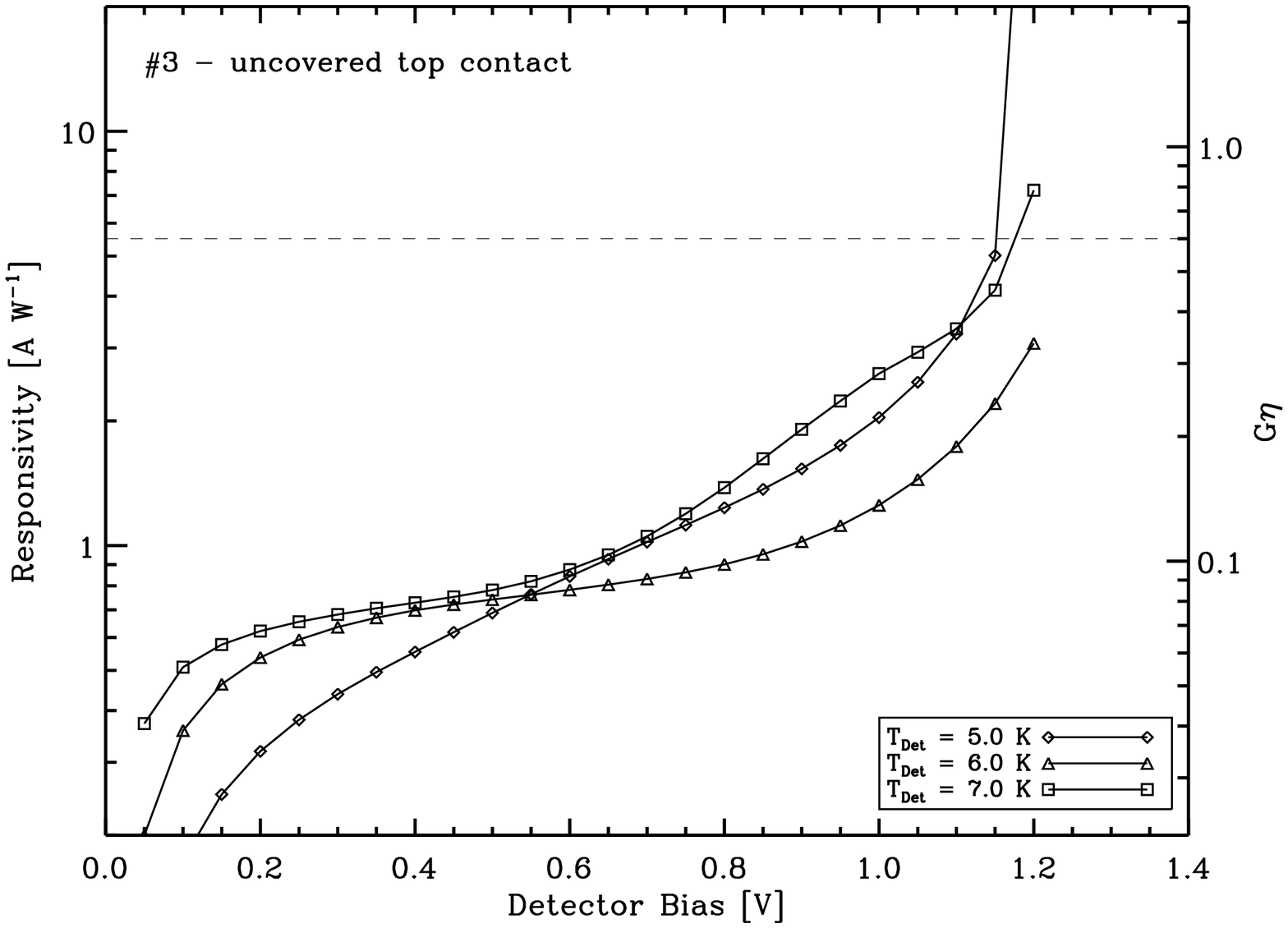}
  \caption{\label{fig:resp}Responsivity (left scale) and the quantum efficiency photoconductive gain product $G\eta$ (right scale) of the three tested detectors for three temperatures and in dependence of the bias voltage, measured at a wavelength of $\lambda = 11.3\,\mathrm{\mu m}$.}
 \end{center}
\end{figure}

The responsivity at a wavelength of $11.3\,\mathrm{\mu m}$ is shown in Fig.~\ref{fig:resp} for the three detectors. The measured values for detectors \#1 and \#2 agree very well and for moderate bias voltages ($U_\mathrm{Bias} \approx 0.8\ldots 1.0$\,V) the responsivity is in the order of $5\,\mathrm{A\,W^{-1}}$. In contrast, the responsivity for \#3 is lower by a factor of $\approx 5$. Furthermore, the bias voltage at which detector breakdown occurs is lower than for the other detectors. We have no obvious explanation for this behavior, given that the dark current densities are comparable for all three devices.

Measuring the quantum efficiency $\eta$ imposes another difficulty, because the photocurrent $I_\mathrm{Phot}$ also depends on the photoconductive gain $G$:
\begin{equation} \label{eq:phcurr}
 I_\mathrm{Phot} = Q A_\mathrm{Det} e G \eta,
\end{equation}
where $e$ is the elemental charge. Therefore by measuring the photocurrent $I_\mathrm{Phot} = I_\mathrm{Det,open} - I_\mathrm{Det,closed}$ one only gets the product of $G \eta$, with both factors being bias dependent. The results are also presented in Fig.~\ref{fig:resp}.

In general it is also possible to determine $G$ or at least the gain dispersion product $\beta G$ by measuring the detector noise for different signal levels\cite{galdemard:129}. However, a prerequisite for this method is that the measured detector noise is dominated by the photon noise and that a significant flux range is covered. With our setup the range in photon flux is limited by the warm black body and thus the dynamic range is not large enough to determine $\beta G$. On the other hand experiences with block-impurity-band detectors suggest that on the flat part of the response curve (i.e.\ for $U_\mathrm{Bias} \leq 1.0\,\mathrm{V}$), the photoconductive gain is in general $G \leq 1$\cite{2002dlus.book.....R,galdemard:129}. Under this assumption the values displayed in Fig.~\ref{fig:resp} would reflect the actual quantum efficiency of the detectors at $\lambda = 11.3\,\mathrm{\mu m}$.

\subsection{Relative Spectral Response}

The relative spectral response is measured by feeding the light from a dual grating monochromator to the detector and a bolometer using infrared fibers. The bolometer is used as a reference to correct for the grating efficiency, the transmission of the order sorting filters, and the unknown and changing transparency of the atmosphere in the lightpath. The relative response is determined by dividing the output singal of the detector by that of the bolometer and normalizing the result to one.

\begin{figure}[p]
 \begin{center}
  \includegraphics[height=0.3\textheight]{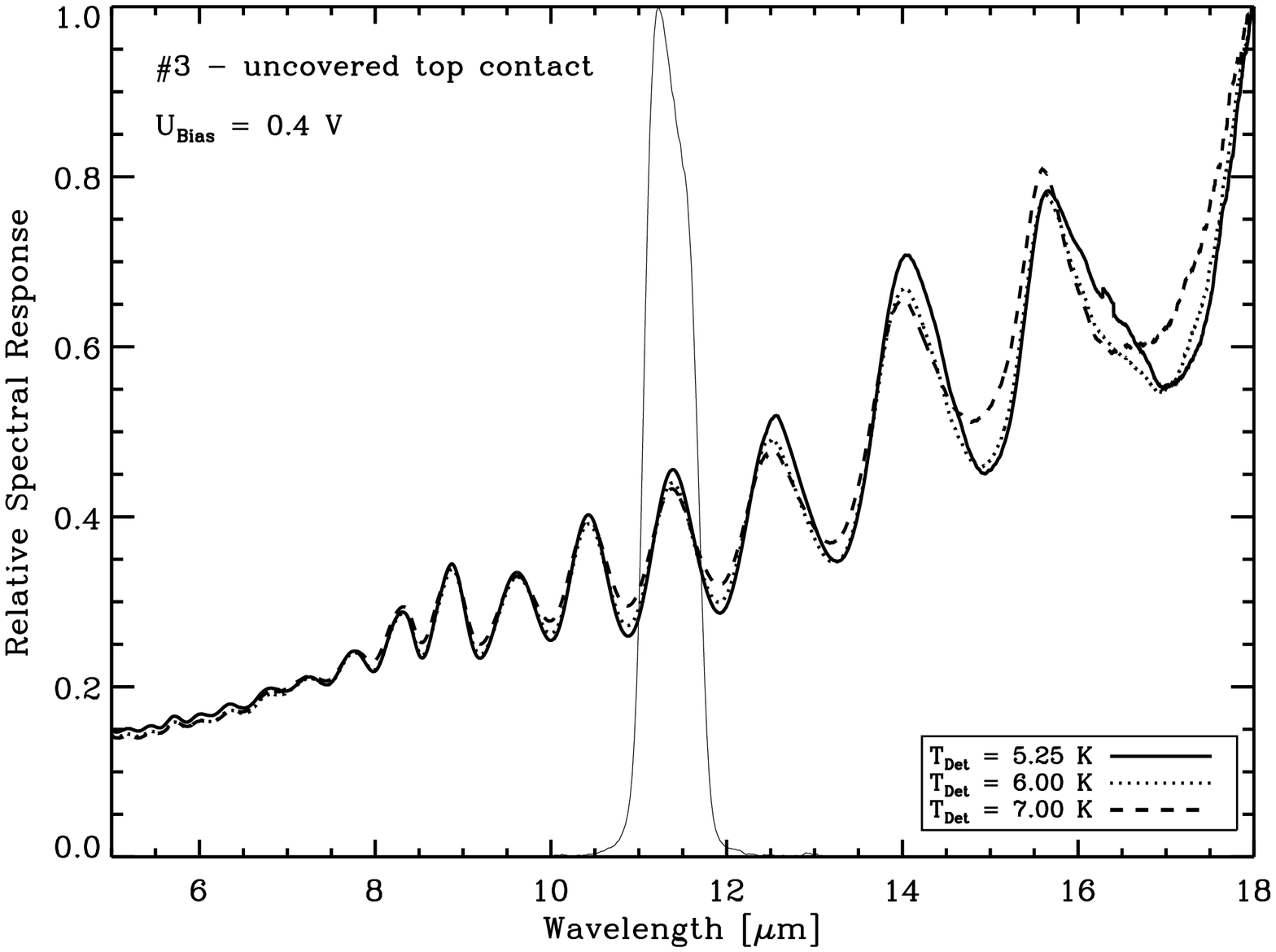}
  \includegraphics[height=0.3\textheight]{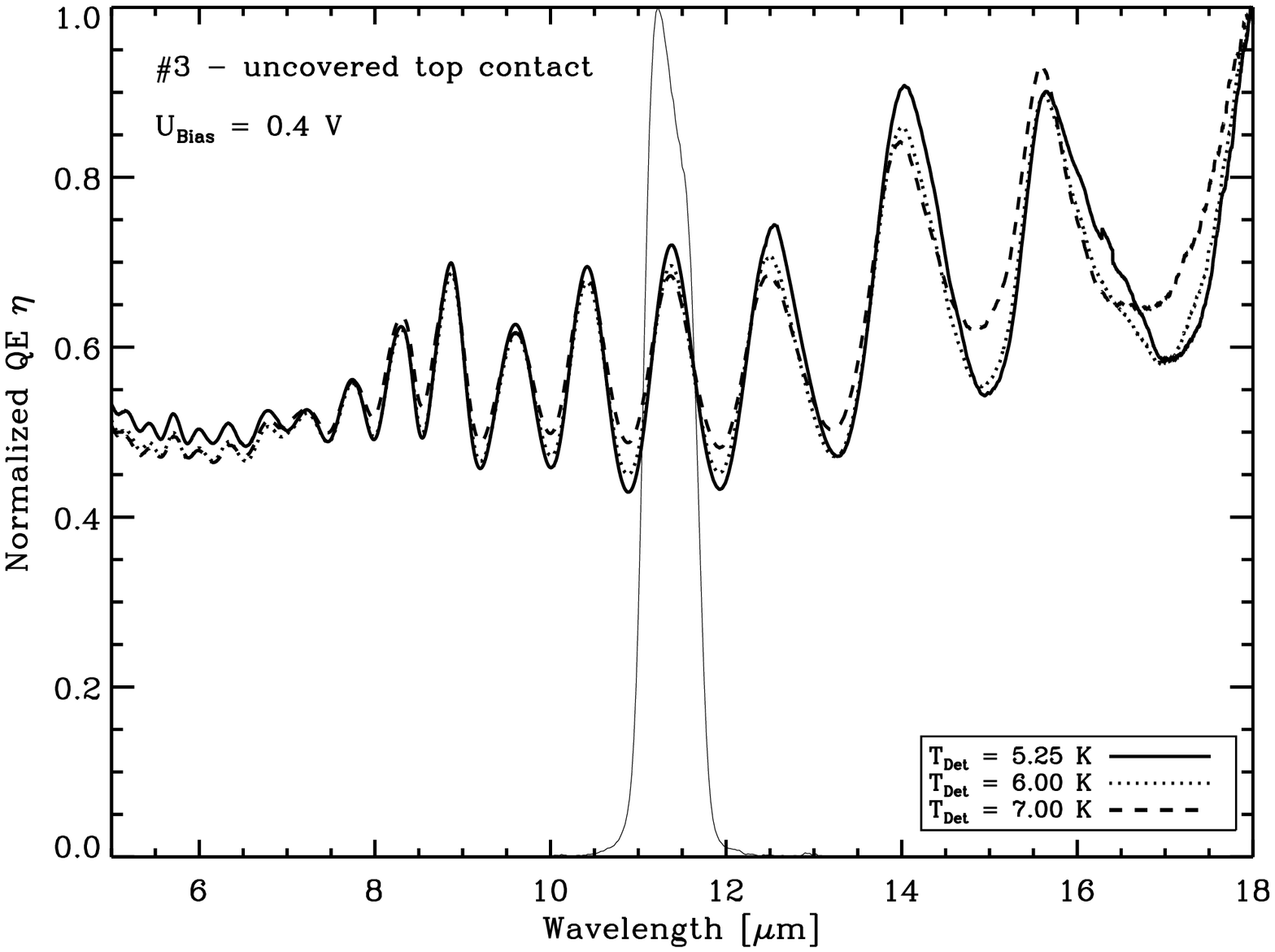} 
  \includegraphics[height=0.3\textheight]{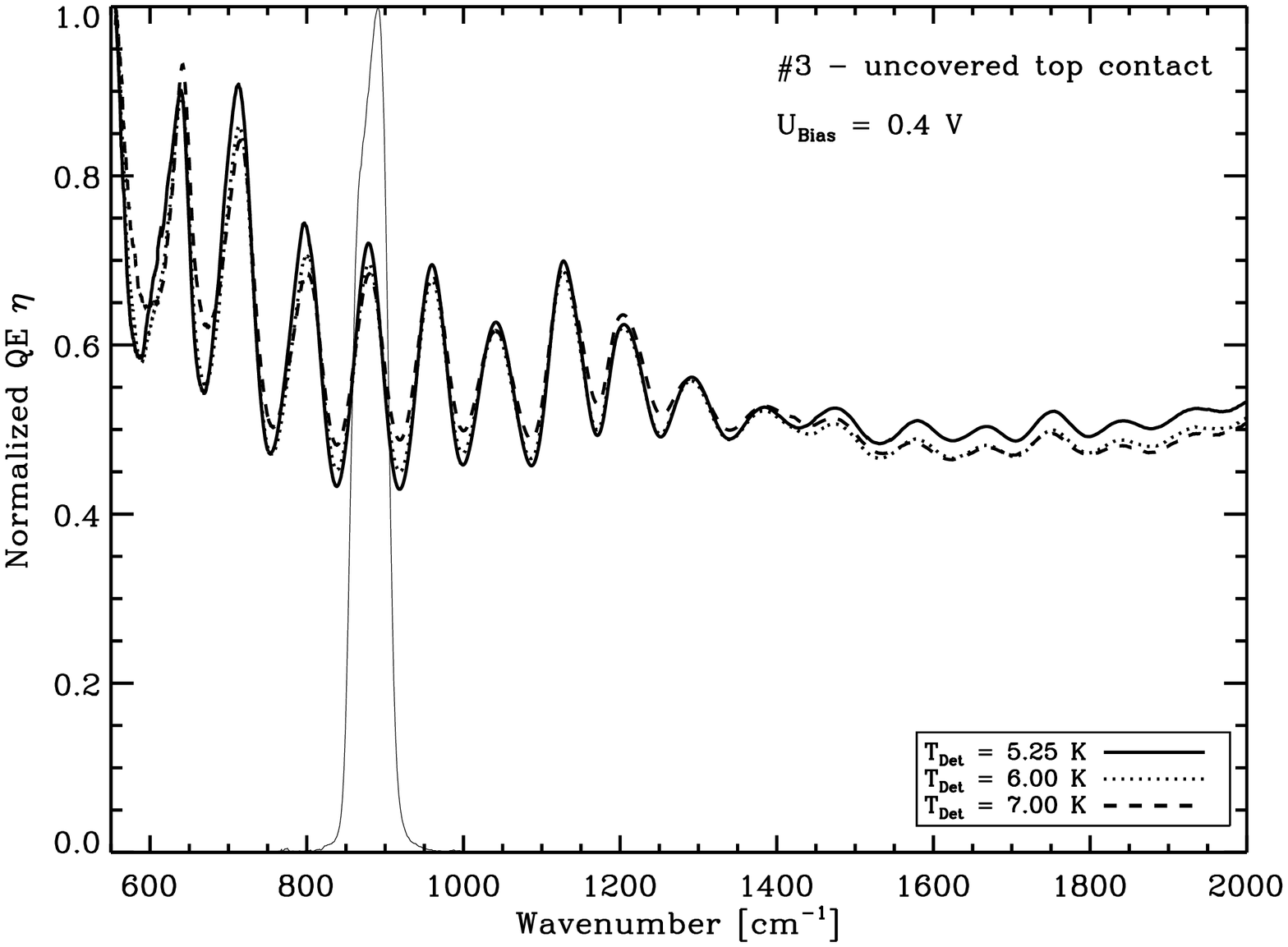}
  \caption{\label{fig:relspec} The relative spectral response of detector \#3 for different detector temperatures. Please note that the upper panel shows the response per incoming flux in units power, whereas the middle panel displays the relative response per photon, thus showing the wavelength dependence of the quantum efficiency $\eta$. The bottom panel shows the same data as the middle panel, but plotted over the wavenumber. In all panels the thin line denotes the normalized transmission of the bandpass filter used to measure the response and quantum efficiency in section~\ref{sec:resp}.}
 \end{center}
\end{figure}

Fig.~\ref{fig:relspec} displays the results for detector \#3 for the wavelength range from $5\,\mathrm{\mu m}$ to $18\,\mathrm{\mu m}$. There seems to be little dependence on the detector temperature $T_\mathrm{Det}$, the same is observed for the bias voltage $U_\mathrm{Bias}$ (not shown). The pronounced oscillations in the response especially for wavelengths $\lambda \geq 8\,\mathrm{\mu m}$ are most probably due to multiple reflections and interference in the detector. When plotting the spectral response over the wavenumber the peaks are equidistant with a separation of about $80\,\mathrm{cm^{-1}}$, supporting this explanation (see Fig.~\ref{fig:relspec}, bottom panel).
The first suspect for this interference is the uncoated silicon substrate. However, the separation of the peaks in the response is too wide given the thickness of the substrate is $d\approx 600\,\mathrm{\mu m}$. According to etalon theory\cite{1986fpi..book.....H}, the wavelength separation between adjacent transmission peaks $\delta \lambda$ is given by
\begin{equation} \label{eq:etalon}
 \delta \lambda = \frac{\lambda_0^2}{2n d\cos\theta},
\end{equation}
where $\lambda_0$ is the wavelength of the nearest peak, $n$ the wavelength and temperature dependent index of refraction of silicon, and $\theta$ the angle the light travels through the substrate. In our case the angle is close to normal incidence $\cos\theta \approx 1$ and the index of refraction of silicon at low temperatures is $n \approx 3.38$ for $\lambda \geq 6\,\mathrm{\mu m}$\footnote[2]{see http://www.irfilters.reading.ac.uk/library/technical\_data/infrared\_materials/si3.htm}. Therefore the thickness of the interfering layer can be determined by
\begin{equation}
 d = \left(\frac{\lambda_0^2}{\delta\lambda}-\lambda_0\right) / \left(2n\right).
\end{equation}
With the data presented in Fig.~\ref{fig:relspec} we find $d\approx 15\,\mathrm{\mu m}$, which matches the thickness of the active layer of the detectors\cite{tezcan:66600R}.

\subsection{Effects of $\gamma$-Irradiation}

We have irradiated a test device using a $^{137}$Cs-$\gamma$-source (activity $A \approx 1.56\,\mathrm{GBq}$, $E_\gamma = 662\,\mathrm{keV}$) with an effective dose rate of approximately $10\,\mathrm{mGy\,h^{-1}}$. Before, during, and after the irradiation the detector signal and noise was recorded. In addition, we also searched for glitches in the output signal using a digital oscilloscope. During all parts of the test the detector was illuminated with a flux density of $F \approx 2\times 10^{-6}\,\mathrm{W\, cm^{-2}}$.

Despite the comparably high dose rate during irradiation there was no obvious change in neither the responsivity nor the noise observed. This might be partly due to the relatively large flux onto the detector, possibly drowning the glitches and leading to a strong self-curing, not allowing the responsivity to increase significantly. Therefore this test will be repeated under dark conditions to recheck.

\section{SUMMARY} 

We tested three Si:As blocked-impurity-band detector prototypes developed by IMEC in the framework of the \textsc{Darwin} mission. All devices are operational and show response to infrared light. We measured the dark current, responsivity, quantum efficiency, and relative spectral response for different detector temperatures and bias voltages. While the dark current density is comparable for all devices, the responsivity of one detector (\#3) is significantly reduced with respect to the others. The results for detectors \#1 and \#2 are summarized in table~\ref{tab:summ}.

\begin{table}[htbp]
 \begin{center}
  \caption{\label{tab:summ}The key characteristics of the test detectors \#1 and \#2 for a bias voltage of $U_\mathrm{Bias}=0.8\,\mathrm{V}$.}
  \begin{tabular}{l l l}\hline\hline\\[-6pt]
   \textbf{Parameter}&\multicolumn{2}{c}{\textbf{Value}}\\[4pt]\hline
   & \\[-6pt]
   Dark current density&$10^{-11}\ldots 10^{-10}\,\mathrm{A\,cm^{-2}}$&at $T_\mathrm{Det} = 5.0\,\mathrm{K}$\\
   &$10^{-8}\ldots 10^{-7}\,\mathrm{A\,cm^{-2}}$&at $T_\mathrm{Det} = 7.0\,\mathrm{K}$\\[4pt]
   Responsivity&$\approx 5\,\mathrm{A\,W^{-1}}$&at $\lambda = 11.3\,\mathrm{\mu m}$\\
   &$\approx 11\,\mathrm{A\,W^{-1}}$&at $\lambda = 18\,\mathrm{\mu m}$\\
   &$\approx 16\,\mathrm{A\,W^{-1}}$&at $\lambda = 26\,\mathrm{\mu m}$ (estimate)\\[4pt]
   Quantum efficiency&$\approx 45\%$&at $\lambda = 11.3\,\mathrm{\mu m}$\\
   &$\approx 60\%$&at $\lambda = 18\,\mathrm{\mu m}$\\[4pt]
   Specific detectivity $D^*$&$\approx 1\times 10^{13}\,\mathrm{cm\,Hz^{1/2}\,W^{-1}}$&for $T_\mathrm{Det}\leq 7\,\mathrm{K}$\\
   &\multicolumn{2}{l}{(limited by photon noise)}\\[4pt]\hline\hline
  \end{tabular}
 \end{center}
\end{table}

 
\appendix    



\bibliography{article.bbl}   

\end{document}